\documentclass{article}

\usepackage[utf8]{inputenc}
\usepackage[english]{babel} 
\usepackage{graphicx}          
\usepackage{cite}              
\usepackage{color}
\usepackage{amsmath}

\begin{document}

%\fontsize{15}{18}
\selectfont

\begin{center}
\textbf{Analysis of Numerical Algorithms for Computing Rapid Momentum Transfers between the Gas and Dust in Simulations of Circumstellar Disks}
\end{center}

\begin{center}
O.P.~Stoyanovskaya\footnote{Boreskov Institute of Catalysis, Lavrentieva, 5, 630090, Novosibirsk, Russia, stop@catalysis.ru}, E.I.~Vorobyov\footnote{Institute of Astrophysics, University of Vienna, Vienna, Austria, eduard.vorobiev@univie.ac.at}, V.N.~Snytnikov\footnote{Boreskov Institute of Catalysis, Lavrentieva, 5, 630090, Novosibirsk, Russia, snyt@catalysis.ru} 
\end{center}

\begin{abstract}
Approaches used in modern numerical simulations of the dynamics of dust and gas in circumstellar disks are tested. The gas and dust are treated like interpenetrating continuous media that can exchange momentum. A stiff coupling between the gas and dust phases is typical for such disks, with the dust stopping time much less than the characteristic dynamical time scale. This imposes high demands on the methods used to simulate the dust dynamics. A grid, piecewise-parabolic method is used as the basic algorithm for solving the gas-dynamical equations. Numerical solutions obtained using various methods to compute the momentum exchanges are presented for the case of monodisperse dust.  Numerical solutions are obtained for shock tube problem and the propagation of sound waves in a gas-dust medium.  The studied methods are compared in terms of their ability to model media with (a) an arbitrary (short or long) dust stopping time, and (b) an arbitrary dust concentration in the gas (varying the dust to gas mass ratio from 0.01 to 1). A method for computing the momentum exchange with infinite-order accuracy in time is identified, which makes it possible to satisfy the conditions (a) and (b) with minimal computational costs. A first-order method that shows similar results in the test computations is also presented. It is shown that the proposed first-order method for monodisperse dust can be extended to a regime when the dust is polydisperse; i.e., a regime represented by several fractions with different stopping times. Formulas for computing the gas and dust velocities for polydisperse dust with each fraction exchanging momentum with the gas are presented.

\end{abstract}

\section{Introduction}

Simulating the dynamics of gas–dust circumstellar disks is a topical problem in modern computational astrophysics (see, e.g., \cite{HaworthEtAl2016}). The dust and solid bodies in a circumstellar disk are represented by objects with various sizes, from submicron dust particles to planetary cores. For dust particles whose sizes are less than the mean free path of the gas molecules in the circumstellar disk, the ratio of the typical dust stopping time to the local gas velocity (the so-called Epstein\cite{Weidenschilling1977} regime for gas flowing around a rigid body) is given by \footnote{The commonly used term for this ratio, the ``stopping time'', can lead to confusion, since, in the approximation considered here, a particle does not acquire the velocity of the gas, but instead a constant velocity relative to the gas velocity.}:
\begin{equation}
\label{eq:t_stop_rho}
t_{\rm stop}=\frac{s \rho_{\rm s}}{c_{\rm s} \rho_{\rm g}},
\end{equation}
where $s$ is the radius of a dust particle (assumed to be spherical), $\rho_{\rm s}$ the density (intrinsic density) of the dust material, $\rho_{\rm g}$ the density of the gas in the circumstellar disk, and $c_{\rm s}$ the sound speed in the gas. 

For dust particles with sizes of about 1~$\mu$m the dust stopping time $t_{\rm stop}$ in the disk is of order 100~s (see, e.g., \cite{StoyanovskayaDust,LaibePrice2011Test}), while the dynamics of the disk require simulations covering $10^{11}$~s ($\approx 10^4$~yrs) or more. This means that numerical solutions of the non-stationary equations for a multiphase gas–dust medium with the application of explicit integration schemes lead to unacceptably high computational costs. This raises the question of searching for numerical methods enabling exact integration of the dust trajectories with a time step $\tau$ determined purely by the Courant condition for the gas-dynamical part of the system. A number of approaches have been used to solve this problem (see, e.g., \cite{BaiStone2010ApJS,ZhuDust,ChaNayakshinDust2011,RiceEtAl2004,FranceDustCode,Pignatale2016}), which were systematically analyzed in \cite{StoyanovskayaDust}. 

In addition to the difficulties associated with taking into account the influence of the gas on the dust when integrating the equations for the dust, the correct computation of momentum transfer from the dust to the gas is also problematic (see, e.g., \cite{BateDust2014,LaibePrice2014OneFluidDust,Ishiki2017,YangJohansen2016}). Averaged over the disk, the ratio of the volume density of dust to the gas density does not exceed $\varepsilon=\displaystyle \frac{\rho_{\rm d}}{\rho_{\rm g}}=0.01$, and the influence of the dust on the gas dynamics is therefore often neglected. On the other hand, the results of simulations show that dust particles can be concentrated in certain areas in the disk (e.g., in spiral arms \cite{RiceEtAl2004}, the inner part of the disk \cite{VorobyovEtAl2017}, or self-gravitating gaseous clumps \cite{ChaNayakshinDust2011}), enhancing the local dust-to-gas mass ratio to values of $\varepsilon\approx 1$ or more. 

Laibe and Price \cite{LaibePrice2011Test} solved test problem of the propagation of sound waves in a two-phase medium applying smoothed-particle hydrodynamics. In their numerical model, the gas and dust were described as separate groups of model particles. They found that, in the case of a high drag coefficient between the gas and dust and with a high concentration of dust in the gas (i.e., when $\varepsilon\approx1$), correct computation of the perturbation amplitude
requires that
\begin{equation}
\label{eq:sparesSPH}
h<c_{\rm s} t_{\rm stop}.
\end{equation} 
This condition probably arises in smoothed-particle computations involving convective transport that exceeds the decay of the wave.

Vorobyov et al. \cite{VorobyovEtAl2017} solved test problems of the propagation of sound waves and the shock tube problem but applying the finite-difference, finite-volume grid method described in detail in \cite{StoneNorman1992} to model the dynamics of the gas and dust components of the disk. The semi-implicit scheme presented by Cha and Nayakshin \cite{ChaNayakshinDust2011} was used to compute the mutual drag between the dust and gas. It was established that such computations for a medium with a high drag coefficient between the gas and dust and with a high concentration of dust in the gas encountered problems.  Applying the scheme of \cite{ChaNayakshinDust2011} required that the time step $\tau$ be much smaller than the velocity relaxation time $t_{\rm stop}$. However, when only the influence of the gas on the dust dynamics is taken into account (not the ``back reaction'' of the dust on the gas dynamics), this scheme gives good results for test problems, without a stiff limitation on the time step. 

This raises the question of creating a universal numerical scheme that is free from these limitations on the spatial and temporal steps. The need for a universal numerical scheme is motivated by the following factors. The ``frozen'' solid phase approximation is often used to compute the dynamics of disks with submicron dust (see, e.g., \cite{Drozdovskaya2015, Drozdovskaya2016, DemidovaGrininDust2017}). If it is important to take into account the dust drift, one computationally effective approach to this is a transition to an asymptotic approximation, or to the short-friction-time approximation \cite{JohansenKlahr2005, Akimkin2015SFTA, Akimkin2017SFTA}.In this approximation, the velocities of the bodies and of the gas are related by a simple algebraic expression that yields correct results for the dust drift in circumstellar disks only for grains of a limited size (see \cite{StoyanovskayaDust} or more detail, and see \cite{PriceLaibeSFTA} for smoothed-particle hydrodynamics). On the other hand, the results of the numerical simulations \cite{BrauerDullemondHenning2008} and of observations show that the dust grains in circumstellar disks grow from 1~$\mu$m to 1-10~cm or more over the first 10 million years of the disk’s evolution. Therefore, simulations of the disk dynamics over long time scales impose requirements on the algorithms used, which must enable computation of momentum exchange between the gas and dust for a wide range of sizes for the solid bodies, from 1~$mu$m to tens of meters, with corresponding variations in the frictional force. Furthermore, it is important that these algorithms can be included in models for gas disks that have already been developed, and do not require fundamental changes in the method used to compute the gas dynamics, for example the transition to a conservative form of the equations (the scheme for the equations of a two-phase medium in conservative form developed by Miniati \cite{Miniati2010}).

In our current study, we have analyzed numerical schemes for the computation of the mutual drag between the gas and dust, and present approaches that make it possible to develop fast methods for the computation of this force in circumstellar gas-dust disks. The finite-difference and finite-volume method \cite{StoneNorman1992} with piecewise-parabolic interpolation \cite{ColellaWoodward1984} for the gas-dynamical part of the equations is used as an example. We have shown the possibility of computing the rapid exchange of momentum between the gas and dust in a circumstellar disk for an arbitrary dust concentration. We have also verified the necessity of the condition (\ref{eq:sparesSPH})  when using grid methods to solve the gas-dynamical equations. 

Section \ref{sec:discnummodel} presents a brief description of the numerical method used to solve the gas-dynamical equations, which is used as a basis to test various methods for computing the drag between the gas
and the solid phase. The tested schemes for the computation of the momentum exchange between the gas and moodisperse dust (with first-order and infinite-order approximations in time) are described in Section \ref{sec:schemes1}. Sections \ref{sec:DustyWave} and \ref{sec:DustyShock} describe one-dimensional test problems for a gas-monodisperse dust system, and Sections \ref{sec:NumDustyWave} and \ref{sec:NumDustyShock} present the results of these test computations. In Section \ref{sec:schemesN} we generalize the scheme with first-order approximation in time to the case of polydisperse dust and present direct computational formulas. Our conclusions are presented in Section \ref{sec:Resume}.

\section{Numerical Method for the Solution of the Dynamical Equations of a Gas-Dust Disk}
\label{sec:discnummodel}

The dynamics of the dust and small bodies in a gaseous circumstellar disk can be described using a system of gas-dynamical equations in which the dust pressure is negligible compared to the gas pressure (see, e.g., \cite{Garaud2004,ZiglinaMakalkin2016}). In a large number of models that have been developed, the dynamics of the bodies are calculated separately from the dynamics of the gas (see, e.g., \cite{ZhuDust,ChaNayakshinDust2011,FranceDustCode,Pignatale2016,Lehmann2017,Ishiki2017,Saito2003,BargeDust2017,MirandaDust2017,Surville2016,FuDust2014,Rosotti2016,Pinilla2016,Gonzalez2017}). This is called the two-fluid approach. In a number of cases, the system of equations for the two-phase gas–dust medium can conveniently be written in terms of the density of the gaseous carrier, the mass fraction of the dust relative to the gas, barycentric velocity of the medium, and the relative velocity between the gas and the solid bodies (see, e.g., \cite{LaibePrice2014OneFluidDust,Phantom2017}). A fully Lagrangian approach can be used to describe the dust dynamics, in which the equations for the trajectories of model or test particles are solved (\cite{BaiStone2010ApJS,Zhu2014,YangJohCarrera2017,CrnkovicRubsamen2015,1MNRAS}); the advantages and difficulties of this approach are described in \cite{YangJohansen2016,Miniati2010}.

We considered a two-fluid polytropic model for the medium based on an Eulerian approach, in which the gas and dust exchange momentum, but not thermal energy. In this case, the gas–dust disk can be described using standard continuity equations and the equations for the motions of the gas and dust components:

\begin{equation}
\label{eq:gas}
\displaystyle\frac{\partial \rho_{\rm g}}{\partial t}+\nabla (\rho_{\rm g} v)=0,\ \ \ 
 \rho_{\rm g} \left[\displaystyle\frac{\partial v}{\partial t}+(v \cdot \nabla) v \right]=-\nabla P+ \rho_{\rm g} g - f_{\rm drag},
\end{equation}
\begin{equation}
\label{eq:dust}
\displaystyle\frac{\partial \rho_{\rm d}}{\partial t}+\nabla (\rho_{\rm d} u)=0,\ \ \ 
\rho_{\rm d} \left[\displaystyle\frac{\partial u}{\partial t}+(u \cdot \nabla) u \right]=\rho_{\rm d} g+ f_{\rm drag},
\end{equation}
where $\rho_{\rm g}$ and $\rho_{\rm d}$are the volume densities of the gas and dust, $v$ and $u$ the velocities of the gas and dust, $P$ the gas-kinetic pressure, $g$ the gravitational accelerations acting on the gas and dust, and $f_{\rm drag}$ the drag force between the dust and gas. 

The terms in square brackets in (\ref{eq:gas})-(\ref{eq:dust}) based on the method of operator splitting with respect to physical processes. We solved the continuity equations and equations of motion using the finite-difference, finite-volume method described in detail for the case of a one-phase medium by Stone and Norman \cite{StoneNorman1992}.  The first stage in the operator splitting scheme is computing the advective terms responsible for the transport of mass and momentum:
\begin{equation}
\label{adv:gas}
\displaystyle\frac{\partial \rho_{\rm g}}{\partial t}+\nabla (\rho_{\rm g} v)=0,\ \ \ 
\rho_{\rm g} \left[\displaystyle\frac{\partial v}{\partial t}+(v \cdot \nabla) v \right]=0,
\end{equation}
\begin{equation}
\label{adv:dust}
\displaystyle\frac{\partial \rho_{\rm d}}{\partial t}+\nabla (\rho_{\rm d} u)=0,\ \ \ 
\rho_{\rm d} \left[\displaystyle\frac{\partial u}{\partial t}+(u \cdot \nabla) u \right]=0.
\end{equation}
This was carried out using the piecewise-parabolic method \cite{ColellaWoodward1984}, for which a number of modifications are known, such as those of \cite{Popov2007,Bisikalo}. In the second stage, the influence of friction and gravitation on the motion of the gas and dust components in the circumstellar disk is computed using the updated densities and gas velocities from the first stage:
\begin{equation}
\label{source:gas}
\rho_{\rm g} \frac{\partial v}{\partial t}=-\nabla P+ \rho_{\rm g} g - f_{\rm drag},
\end{equation}
\begin{equation}
\label{source:dust} 
\rho_{\rm d} \frac{\partial u}{\partial t}=\rho_{\rm d} g+ f_{\rm drag}.
\end{equation}

Note that formally Eqs. (\ref{adv:dust}) and (\ref{source:dust}) describe the motion of a cold fluid (gas) with zero temperature. The numerical solution of this system can lead to the development of strong discontinuities in the dust velocity and associated instabilities, especially in the presence of self-gravitation \cite{ZhuDust}. Schemes with artificial or physical viscosity, as well as strong drag between the gas and dust, can suppress the development of instability. In practice, the appearance of discontinuities is usually suppressed by introducing a small dust pressure comprising a few percent (or less) of the gas pressure in the corresponding numerical cells.

We have analyzed various ways of solving equations of the type (\ref{source:gas})-(\ref{source:dust}), and compared these methods in terms of their suitability for problems in which the dust is stiffly coupled to the gas; i.e., the dust stopping time $t_{\rm stop}$ is much shorter than the dynamical time scale of the circumstellar disk. Note that the system (\ref{adv:gas})-(\ref{adv:dust})can be solved using any method to compute the convective terms in the hydrodynamical equations (see, e.g., \cite{vanLeer1977}). We used the piecewise-parabolic method \cite{ColellaWoodward1984}, since it has third-order accuracy in space.

\section{Approaches to Computing the Rapid Exchange of Momentum between
the Gas and Bodies. Tested Schemes }
\label{sec:schemes1}
We considered a medium in which bodies interact with the gas in the Epstein regime (according to \cite{Weidenschilling1977} $s<2.25\lambda$, where $\lambda$ is the mean free path of the gas molecules), that is,
\begin{equation}
\label{eq:fdrag}
f_{\rm drag}=\rho_{\rm d} \displaystyle \frac{v-u}{t_{\rm stop}}.
\end{equation}
A linear relationship between the drag force and the relative velocity between the gas and dust takes place for a wide range of body sizes, due to the low density of the gas in circumstellar disks. The maximum size of dust particles in the disk that can adequately be described using this approximation is determined in \cite{StoyanovskayaDust}, Fig.2. 
Setting $\varepsilon=\displaystyle\frac{\rho_{\rm d}}{\rho_{\rm g}}$, where $\varepsilon$ is the mass fraction of the dust relative to the gas, 
\begin{equation}
a_{\rm g}=-\displaystyle\frac{\nabla P}{\rho_{\rm g}}+g_{\rm g}, \ \ a_{\rm d}=g_{\rm d}.
\end{equation}
and $a_{\rm g}$, $a_{\rm d}$ are the accelerations of the gas and dust
apart from the drag force acceleration, Eqs. (\ref{source:gas})-(\ref{source:dust}) acquire the form
\begin{equation}
\label{eq:system}
\left\{
 \begin{array}{lcl}
        \displaystyle 
        \frac{\partial v}{\partial t} = a_{\rm g}-\varepsilon\displaystyle\frac{v-u}{t_{\rm stop}}, \\
        \displaystyle 
        \frac{\partial u}{\partial t} = a_{\rm d} + \displaystyle\frac{v-u}{t_{\rm stop}}. 
    \end{array}
\right.
\end{equation}

Let the velocity of a specified volume of gas and dust be known at some time,
\begin{equation}
\label{eq:systemInit}
v|_{t=t_0}=v_n, u|_{t=t_0}=u_n,
\end{equation}
We then find the velocities $v^{n+1}$ and $u^{n+1}$ displayed by this same volume of gas and dust after a time interval $\tau$. 

To obtain stable solutions using an explicit, first-order approximation scheme,
\begin{equation}
\label{eq:explicitEuler}
\left\{
 \begin{array}{lcl}
 \displaystyle\frac{v^{n+1}-v^{n}}{\tau}=a_{\rm g}-\varepsilon\displaystyle\frac{v^n-u^n}{t_{\rm stop}},\\
 \displaystyle\frac{u^{n+1}-u^n}{\tau}=a_{\rm d}+\frac{v^n-u^n}{t_{\rm stop}},\\
 \end{array}
\right.
\end{equation}
the eigenvalues of the matrix for the transition from ($v^n$,$u^n$) to ($v^{n+1}$,$u^{n+1}$) must be less than unity in magnitude. For $\Delta=\displaystyle \frac{\tau}{t_{\rm stop}}$, the system (\ref{eq:explicitEuler}) is equivalent to
\begin{equation}
\label{eq:ExplicitMatrix}
\begin{pmatrix}
v^{n+1}\\
u^{n+1}
\end{pmatrix}
=
\begin{pmatrix}
1-\varepsilon\Delta & \varepsilon\Delta \\
\Delta & 1-\Delta
\end{pmatrix}
\begin{pmatrix}
v^n\\
u^n
\end{pmatrix}
+
\begin{pmatrix}
a_{\rm g}\\
a_{\rm d}
\end{pmatrix}.
\end{equation}

The eigenvalues of this matrix are $\lambda_1=1$ and $\lambda_2=1-\Delta(\varepsilon+1)$; i.e., stability requires the use of a time step satisfying the condition
\begin{equation}
\label{eq:tauexplicit}
\tau<\displaystyle \frac {2 t_{\rm stop}}{\varepsilon+1}.
\end{equation}
The following approaches are used to remove the limitations of the explicit scheme for the integration of (\ref{eq:system})-(\ref{eq:systemInit}).

\subsection{Semi-implicit Scheme with Operator Splitting}
This scheme is constructed by analogy with the semi-implicit scheme, or mixed-time-layer scheme, of \cite{StoyanovskayaDust,ChaNayakshinDust2011}, in which stability of the solution is provided by using the aerodynamical velocity drag from the following time layer in the computation.
\begin{equation}
\label{eq:InitFinite}
\left\{
 \begin{array}{lcl}
 \displaystyle\frac{v^{n+1}-v^{n}}{\tau}=a_{\rm g}-\varepsilon\displaystyle\frac{v^{n+1}-u^n}{t_{\rm stop}},\\
 \displaystyle\frac{u^{n+1}-u^n}{\tau}=a_{\rm d}+\frac{v^{n+1}-u^{n+1}}{t_{\rm stop}}.\\
 \end{array}
\right.
\end{equation}
The equations in the system (\ref{eq:InitFinite}) are solved consecutively; that is, the first explicitly yields $v^{n+1}$, then the second yields $u^{n+1}$.  Test computations for this scheme are presented in \cite{VorobyovEtAl2017}. 

When $a_{\rm g}=0$, $a_{\rm d}=0$ and $\varepsilon=const$ it follows
from (\ref{eq:system}) that 
\begin{equation}
\label{eq:conserv}
\displaystyle\frac{\partial}{\partial t}(v+\varepsilon u)=0.
\end{equation}
However, for the numerical solution found using the scheme (\ref{eq:InitFinite}), in place of the discrete analog of relation (\ref{eq:conserv}) $v^{n+1}+\varepsilon u^{n+1}=v^n+\varepsilon u^n$, we have the relation
\begin{equation}
\label{eq:accuracy}
v^{n+1}+\varepsilon u^{n+1}=v^n+\varepsilon u^n+\displaystyle\frac{\tau}{t_{\rm stop}}\varepsilon (u^n-u^{n+1}).
\end{equation}
Thus, if $\tau \gg t_{\rm stop}$ and $a_{\rm g} \ll \displaystyle \varepsilon \frac{v-u}{t_{\rm stop}}$, we expect appreciable deviations of the numerical solution obtained using the scheme (\ref{eq:InitFinite}) from $v + \varepsilon u$, due to the high value of the third term in(\ref{eq:accuracy}). Furthermore, it is clear that the accuracy of the numerical solution will grow as $\varepsilon$ is decreased, due to the decrease in the absolute value of this term. The dependence of the exact solution on $\varepsilon$ coincides with the results of Vorobyov et al. \cite{VorobyovEtAl2017}, and also with the results of Laibe and Price \cite{LaibePrice2011Test}, obtained using smoothed-particle hydrodynamics.

\subsection{Semi-Analytical Scheme}
Because $a_{\rm g}$, $a_{\rm d}$, $t_{\rm stop}$ and $\varepsilon$ are constant at each moment in time, the solution of (\ref{eq:system})-(\ref{eq:systemInit}) can be found analytically: 
\begin{equation}
\label{eq:ExpConst}
E=\displaystyle\frac{C_1+(a_{\rm g}+\varepsilon a_{\rm d})\tau}{\varepsilon + 1}, \ \  
D=\displaystyle\frac{C_2}{\varepsilon+1}e^{-\displaystyle\frac{\varepsilon+1}{t_{\rm stop}}\tau}+
\frac{(a_{\rm g}-a_{\rm d})t_{\rm stop}}{(\varepsilon+1)^2},
\end{equation}

\begin{equation}
\label{eq:ExpInit}
C_1=v^n+\varepsilon u^n, \ \ C_2=v^n-u^n-\displaystyle\frac{(a_{\rm g}-a_{\rm d})t_{\rm stop}}{\varepsilon+1},
\end{equation}

\begin{equation}
\label{eq:ExpAll}
u^{n+1}=-D+E, \ \ v^{n+1}=\varepsilon D + E.
\end{equation}

This approach is used in the circumstellar-disk models \cite{PanPadoan2013,Rosotti2016}, without taking into account the back reaction of the dust on the gas, or \cite{BateDust2015,Ishiki2017}, where the back reaction of the dust is included.  Note that a similar idea for developing  computational schemes is used in magnetohydrodynamics (see, e.g.\cite{Inoue2008}), where the rapid exchange of momentum between neutral gas and the plasma components is computed. 

\subsection{Semi-implicit Scheme for the Relative and Barycentric Velocities}

Here, we consider an approach that preserves the implicit nature of the method while making it possible to obtain an acceptable accuracy for the solutions, even when $\tau \gg t_{\rm stop}$. Let us turn to a system of equations for the relative velocity $x=v-u$ and the barycentric velocity $y=v+\varepsilon u$  of the gas-dust medium, equivalent to the system (\ref{eq:system}):

\begin{equation}
\label{eq:Newsystem}
\left\{
 \begin{array}{lcl}
    
        \displaystyle 
        \frac{\partial x}{\partial t} = (a_{\rm g}-a_{\rm d})-\displaystyle\frac{\varepsilon+1}{t_{\rm stop}}x, \\
        \displaystyle 
        \frac{\partial y}{\partial t} = a_{\rm g}+\varepsilon a_{\rm d}. 
    \end{array}
\right.
\end{equation}

Approximating the first equation using an implicit scheme with first-order accuracy in time and writing the exact solution of the second equation yields

\begin{equation}
\label{eq:Finite}
\left\{
 \begin{array}{lcl}
        x^{n+1} = \displaystyle \frac{x^n+\tau(a_{\rm g}-a_{\rm d})}{1+(\varepsilon+1)\displaystyle\frac{\tau}{t_{\rm stop}}}, \\
        y^{n+1} = y^n+\tau (a_{\rm g}+\varepsilon a_{\rm d}). 
    \end{array}
\right.
\end{equation}

\begin{equation}
\label{eq:FiniteVU}
\left\{
 \begin{array}{lcl}
        v^{n+1} = \displaystyle \frac{\varepsilon x^{n+1}+y^{n+1}}{\varepsilon+1}, \\
        u^{n+1} = \displaystyle \frac{y^{n+1}-x^{n+1}}{\varepsilon+1}. 
    \end{array}
\right.
\end{equation}

These formulas coincide with the fully implicit scheme for (\ref{eq:system}):

\begin{equation}
\label{eq:InitImplicit}
\left\{
 \begin{array}{lcl}
 \displaystyle\frac{v^{n+1}-v^{n}}{\tau}=a_{\rm g}-\varepsilon\displaystyle\frac{v^{n+1}-u^{n+1}}{t_{\rm stop}},\\
 \displaystyle\frac{u^{n+1}-u^n}{\tau}=a_{\rm d}+\frac{v^{n+1}-u^{n+1}}{t_{\rm stop}}.\\
 \end{array}
\right.
\end{equation}

In the following sections, we present numerical solutions of two test problems obtained using the semi-implicit scheme with separation, the semi-implicit scheme for the barycentric and relative velocities, and the semi-analytical scheme.

\section{Test 1. DUSTYWAVE --- Propagation of Sound Wave in a Periodic
Two-Phase Medium}
\subsection{Formulation of the DustyWave Problem}
\label{sec:DustyWave}

Let us consider the system made up of the continuity equations and the equations of motion for a two-phase medium:

\begin{equation}
\label{eq:DustyWaveCont}
\frac{\partial \rho_{\rm g}}{\partial t}+\frac{\partial({\rho_{\rm g} v})}{\partial x} = 0, \ \  
\frac{\partial \rho_{\rm d}}{\partial t}+\frac{\partial{(\rho_{\rm d} u)}}{\partial x} = 0,\ \ 
\end{equation}

\begin{equation}
\label{eq:DustyWaveMotionGas}
\rho_{\rm g} (\frac{\partial v}{\partial t}+v \frac{\partial v}{\partial x}) = - c_s^2 \frac{\partial \rho_{\rm g}}{\partial x} - K(v-u),
\end{equation}

\begin{equation}
\label{eq:DustyWaveMotionDust}
\rho_{\rm d} (\frac{\partial u}{\partial t}+u \frac{\partial u}{\partial x}) =  K(v-u),
\end{equation}
where $K=\displaystyle\frac{\rho_{\rm d}}{t_{\rm stop}}$ is the drag coefficient between the gas and the dust. The stationary solution of the system (\ref{eq:DustyWaveCont})-(\ref{eq:DustyWaveMotionDust}) is given by the functions
\begin{equation}
\label{eq:SteadySolution}
\rho_{\rm g}(x)=\tilde{\rho_{\rm g}}=const, \ \
\rho_{\rm d}(x)=\tilde{\rho_{\rm g}}=const, \ \ 
v(x)=0, \ \ u(x)=0.
\end{equation}

Consider the solution of (\ref{eq:DustyWaveCont})-(\ref{eq:DustyWaveMotionDust}) in the interval $x \in [0,1]$ with a positive sound speed, specifying for the solutions at the left-hand boundary of the periodic values of the functions of $x$: 
\begin{equation}
\rho_{\rm g}|_{x=0}=\rho_{\rm g}|_{x=1}, \ \ \rho_{\rm d}|_{x=0}=\rho_{\rm d}|_{x=1}, \ \ v|_{x=0}=v|_{x=1}, \ \ u|_{x=0}=u|_{x=1},
\end{equation}
and initial data in the form of small perturbations of the stationary density and velocity (\ref{eq:SteadySolution}):
\begin{equation}
\label{eq:DustyWave_init1}
\rho_{\rm g}|_{t=0}=\tilde{\rho_{\rm g}}+ A \sin(kx), \ \rho_{\rm d}|_{t=0}=\tilde{\rho_{\rm d}}+ A \sin(kx), 
\end{equation}
\begin{equation}
\label{eq:DustyWave_init2}
v|_{t=0}= A \sin(kx), \ u|_{t=0}= A \sin(kx).
\end{equation}
Here, $k$ is the wave number specifying the integer number of sinusoidal waves of the density and velocity in the interval $[0,1]$, and $A$ is the perturbation amplitude. In the vicinity of (\ref{eq:SteadySolution}) the linearized system (\ref{eq:DustyWaveCont})-(\ref{eq:DustyWaveMotionDust}) will have the form

\begin{equation}
\label{eq:LinDustyWaveCont}
\frac{\partial \delta \rho_{\rm g}}{\partial t}+\tilde{\rho_{\rm g}} \frac{\partial{v}}{\partial x} = 0, \ \  
\frac{\partial \delta \rho_{\rm d}}{\partial t}+\tilde{\rho_{\rm d}} \frac{\partial{ u}}{\partial x} = 0,\ \ 
\end{equation}

\begin{equation}
\label{eq:LinDustyWaveMotionGas}
\tilde {\rho_{\rm g}} \frac{\partial v}{\partial t} = - c_s^2 \frac{\partial \delta \rho_{\rm g}}{\partial x} - K(v-u),
\end{equation}

\begin{equation}
\label{eq:LinDustyWaveMotionDust}
\tilde {\rho_{\rm d}} \frac{\partial u}{\partial t} =  K(v-u).
\end{equation}

The analytical solution of the linearized system (\ref{eq:LinDustyWaveCont})-(\ref{eq:LinDustyWaveMotionDust}) is presented by Laibe and Price \cite{LaibePrice2011}, who also made available their code for the automated generation of this solution that we used. 

Further, for simplicity, we will call the analytical solution of the system (\ref{eq:DustyWaveCont})-(\ref{eq:DustyWaveMotionDust}) the exact solution of the linearized system (\ref{eq:LinDustyWaveCont})-(\ref{eq:LinDustyWaveMotionDust}). 

The linearized system (\ref{eq:LinDustyWaveCont})-(\ref{eq:LinDustyWaveMotionDust}) has an analytical solution for suspensions with both small dust granules (a high drag coefficient) and large bodies (a low drag coefficient). This key property of the problem makes it possible to use it to test the universality of numerical schemes; that is, this problem can be used to evaluate the suitability of a method for simulation dynamics of gas and solids of arbitrary size.

\subsection{Numerical Solution of the DustyWave Problem}
\label{sec:NumDustyWave}

\begin{figure*}
\center
  \includegraphics[scale=0.2]{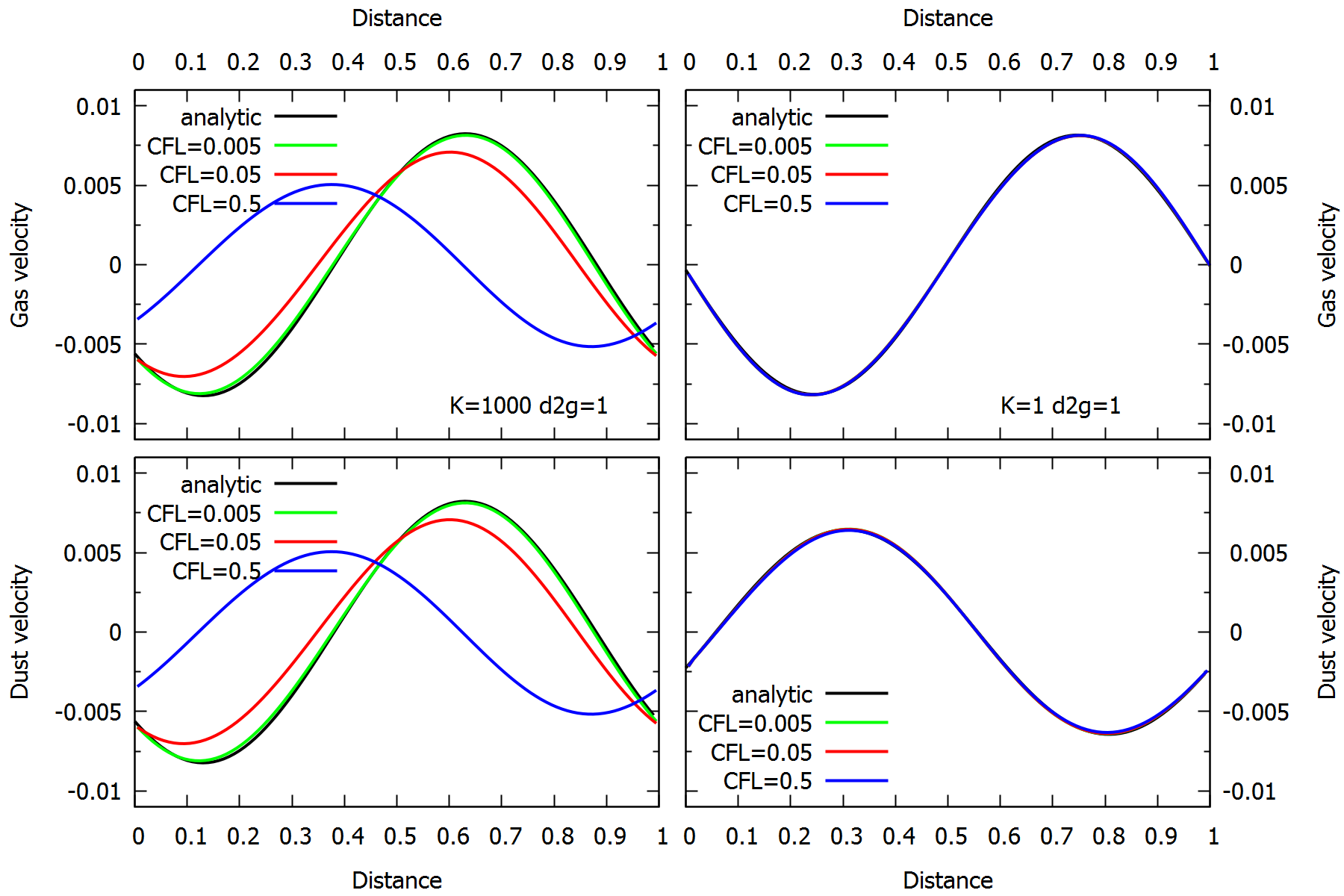} 
  \caption{Solution of (\ref{eq:DustyWaveCont})-(\ref{eq:DustyWaveMotionDust}) at time $t=0.5$ obtained using the semi-implicit scheme with operator splitting (\ref{eq:InitFinite}). The upper panels show the gas velocity and the lower panels the dust velocity. Results for a stiffly coupled medium (drag coefficient $K = 1000$) are shown on the left, and those for a weakly coupled medium (drag coefficient $K=1$) on the right. The black curves shows the analytical solution and the colored curves the numerical solutions for various $CFL$ values.}
\label{fig:InitFinite}
\end{figure*}

In this section, we compare the numerical solutions of the DustyWave problem obtained using the three different schemes. Because the most difficulty is presented by computations of a gas–dust medium with a high dust content (according to the results of \cite{LaibePrice2011Test, VorobyovEtAl2017}), we present here computations for the following parameters:

\begin{equation}
c_s=1, \ \tilde{\rho_{\rm g}}=1, \ \tilde{\rho_{\rm d}}=1, \ k=1, \ A=0.01.
\end{equation}

We used two values of the drag coefficient: $K=1$ (a weakly coupled medium) and $K=1000$ (a stiffly coupled medium). All the computations were conducted on a grid with 100 cells in the interval $[0,1]$ with the artificial viscosity parameter $C2=2$, which assumes smoothing of the solution over two grid cells. We varied the Courant-Friedrichs-Lewy parameter $CFL$ determining the time step in the computations:
\begin{equation}
    \label{eq:tauCFL}
        \displaystyle \tau = CFL \frac{h}{max(c_s, v, u)}.
\end{equation}
We adopted the standard value for the Courant-Friedrichs-Lewy parameter $CFL=0.5$. To understand the properties of the schemes, we reduced the
time step $\tau$ by factors of 10 and 100 relative to the standard value.  

Figure \ref{fig:InitFinite} presents the gas and dust velocities at time $t=0.5$ obtained using the scheme (\ref{eq:InitFinite}). The black curves show the analytical solution and the colored curves the numerical solutions for various $CFL$ values. The gas and dust velocities for a medium with a high drag coefficient are very similar, and the oscillation phases coincide, while the gas and dust velocities and the oscillation phases become considerably different in a medium with a low drag coefficient. On the other hand, the scheme (\ref{eq:InitFinite}) yields results with acceptable accuracy only for $CFL=0.005$. Under the  conditions considered, this $CFL$ value corresponds to $\tau=0.05t_{\rm stop}$. When the time step is increased to $\tau=0.5t_{\rm stop}$ and $\tau=5t_{\rm stop}$, a decrease in the oscillation amplitude is observed and the numerical solution is shifted in phase relative to the analytical solution. On the contrary, there are no appreciable distortions in the solution when the CFL value is varied for a medium with a low drag coefficient.

The effect of the ``numerical'' decay of the oscillations in a stiffly coupled gas-dust medium is described in \cite{LaibePrice2011Test} in relation to another scheme for computing the gas dynamics --- smoothed-particle hydrodynamics. Laibe and Price \cite{LaibePrice2011Test} note that achieving acceptable accuracy requires that the spatial solution satisfies the condition (\ref{eq:sparesSPH}), while reducing the time step without increasing the spatial resolution does not lead to the required accuracy. The condition (\ref{eq:sparesSPH}) is violated with our adopted grid computational method ($h=10 c_s t_{\rm stop}$), but it is possible to closely approach the exact solution by reducing the time step.

\begin{figure*}
\center
  \includegraphics[scale=0.2]{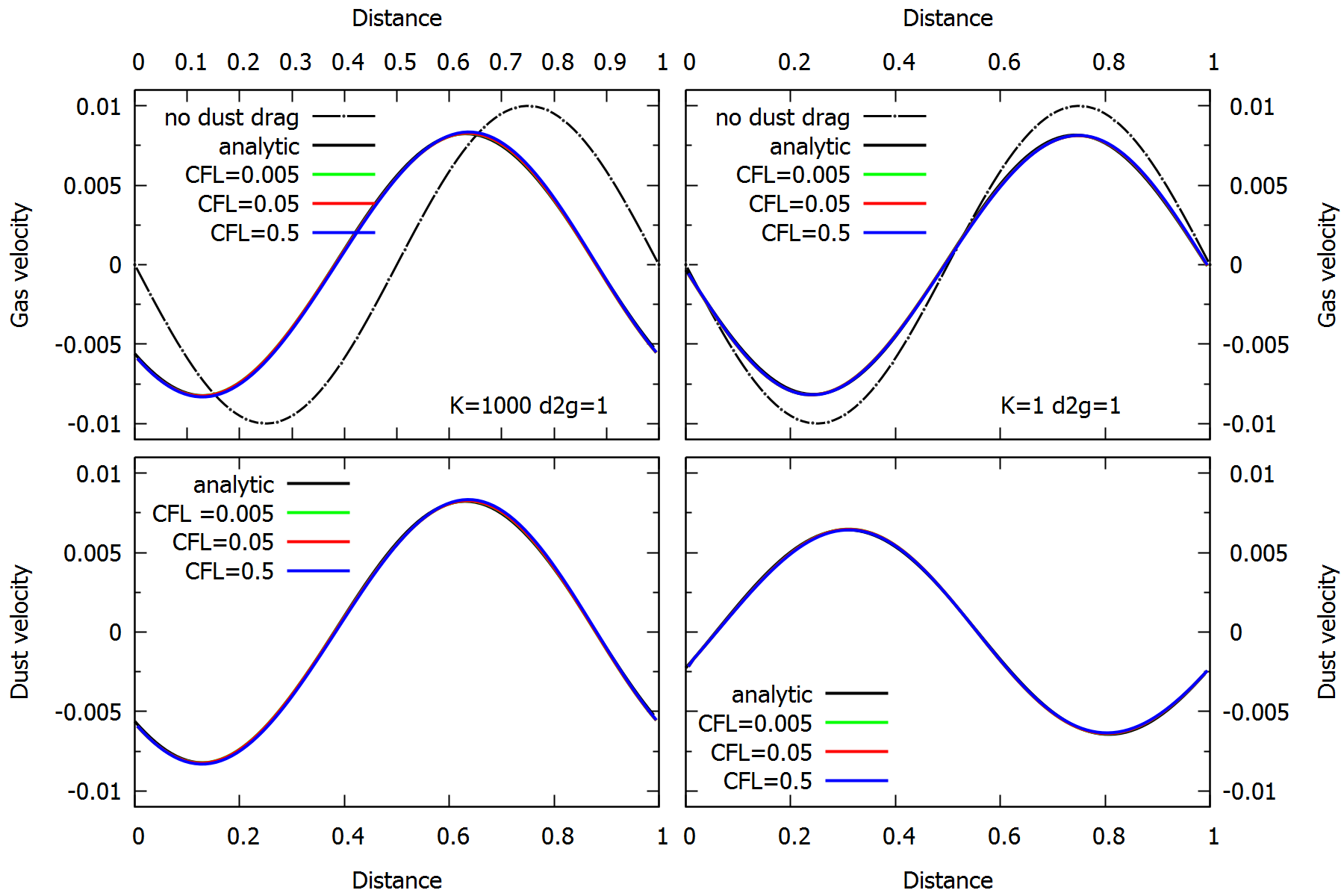} 
  \caption{Solution of (\ref{eq:DustyWaveCont})-(\ref{eq:DustyWaveMotionDust}) at time $t=0.5$ obtained using the semi-analytical scheme (\ref{eq:ExpConst})-(\ref{eq:ExpAll}). The upper panels show the gas velocity and the lower panels the dust velocity. Results for a stiffly coupled medium (drag coefficient $K = 1000$) are shown on the left, and those for a weakly coupled medium (drag coefficient $K=1$) on the right. The colored curves showing the numerical solutions for various $CFL$ values coincide in the figure.}
\label{fig:Exp}
\end{figure*}

Figure \ref{fig:Exp} presents the analogous results for the scheme (\ref{eq:ExpConst})-(\ref{eq:ExpAll}) for various $CFL$ values. The dot-dashed curves in the upper panels show the gas velocity that the medium would have if the dust exerted no influence on the gas dynamics. In a weakly coupled medium, the gas perturbation is transported at a velocity close to the sound speed in the gas, with the presence of dust leading to a decay in the perturbation amplitude. In a stiffly coupled medium with a high dust content, the perturbation is transported at the sound speed in the gas-dust medium $c^*_s$ which is a factor of $\sqrt{2}$ lower than the sound speed in the gas $c_s$ (see (\ref{eq:dustysound}) below). Figure \ref{fig:Finite} presents the results obtained using the scheme (\ref{eq:Finite})-(\ref{eq:FiniteVU}). The propagation of the sound wave is obtained with acceptable accuracy for both a stiffly coupled and weakly coupled medium. Thus, the condition (\ref{eq:sparesSPH}) is not necessary for the piecewise-parabolic advection (PPA) method applied in conjunction with the scheme (\ref{eq:ExpConst})-(\ref{eq:ExpAll}) or (\ref{eq:Finite})-(\ref{eq:FiniteVU}). It is clear that both of the approaches (\ref{eq:ExpConst})-(\ref{eq:ExpAll}) and (\ref{eq:Finite})-(\ref{eq:FiniteVU}) enable the use of a time step determined from the Courant condition (\ref{eq:tauCFL}) for a two-phase medium without additional reduction (\ref{eq:tauexplicit}) due to the appearance of the ``short'' time $t_{\rm stop}$. 

The computations also demonstrated that all the schemes (\ref{eq:InitFinite}), (\ref{eq:ExpConst})-(\ref{eq:ExpAll}), (\ref{eq:Finite})-(\ref{eq:FiniteVU}) yield acceptable computational accuracy when $CFL=0.5$ if the dust content in the gas is sufficiently low, i.e., if $\varepsilon=0.01$.  

\begin{figure*}
\center
  \includegraphics[scale=0.2]{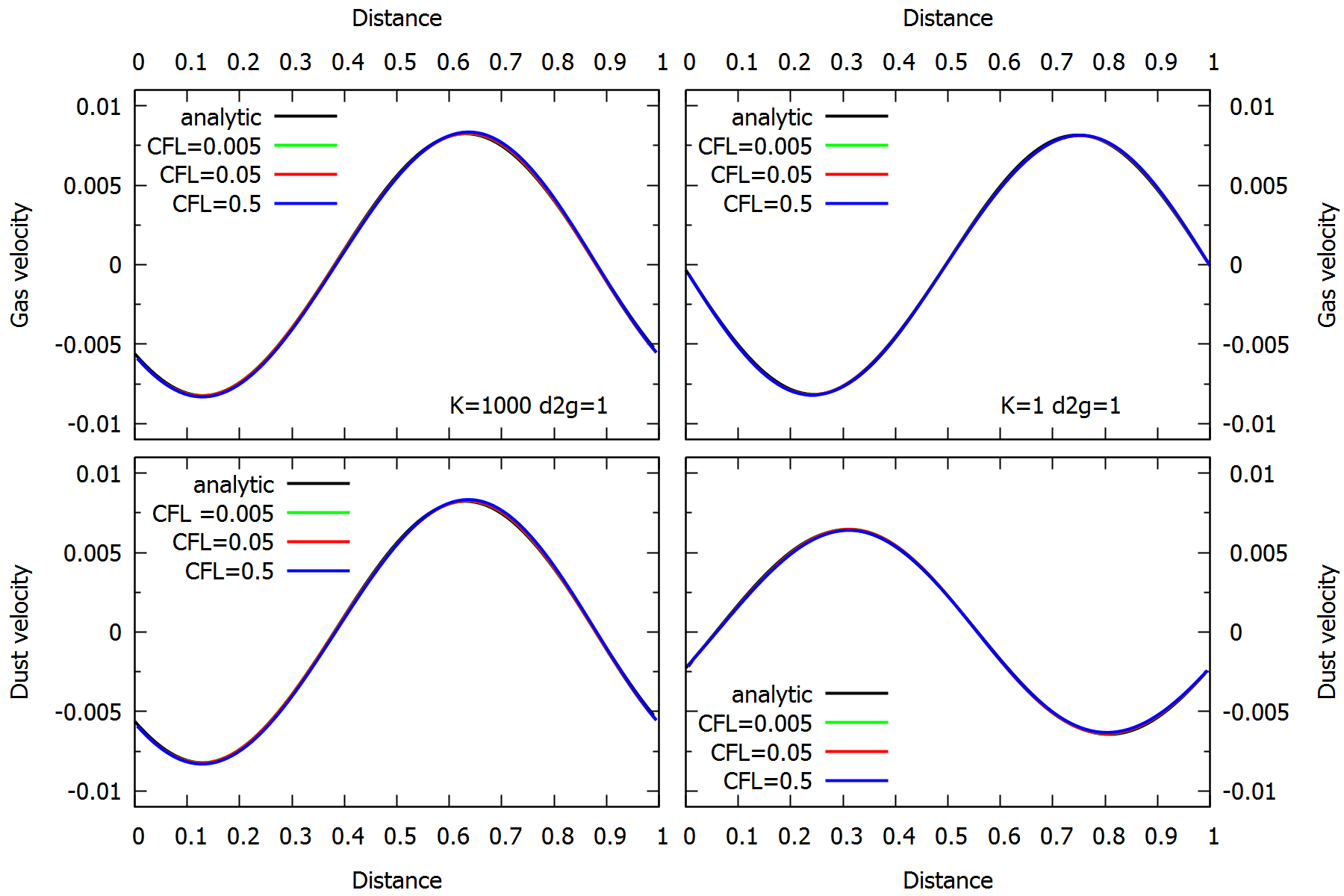} 
  \caption{Solution of (\ref{eq:DustyWaveCont})-(\ref{eq:DustyWaveMotionDust}) at time $t=0.5$ obtained using the barycentric and relative velocity scheme (\ref{eq:Finite})-(\ref{eq:FiniteVU}). The upper panels show the gas velocity and the lower panels the dust velocity. Results for a stiffly coupled medium (drag coefficient $K=1000$) are shown on the left, and those for a weakly coupled medium (drag coefficient $K = 1$) on the right. The colored curves showing the numerical solutions for various $CFL$ values coincide in the figure.}
\label{fig:Finite}
\end{figure*}

\section{Test 2. DUSTYSHOCK - Shock Tube Problem for Gas-Dust Medium}
\subsection{Formulation of the DustyShock Problem}
\label{sec:DustyShock}
Here, we consider the shock tube problem --- a classical test for methods designed for the numerical integration of the dynamical equations of a continuous medium, often referred to as the test of Sod \cite{Sod1978}. This problem has been widely used to test computational schemes for a two-phase medium (see, e.g., \cite{LaibePrice2011Test,ChertokKurganov2017,Saito2003}).
The one-dimensional equations for the conservation of mass, momentum, and energy in a gas-dust medium in the notation of Sections \ref{sec:schemes1} and \ref{sec:DustyWave} have the form

\begin{equation}
\label{eq:ShockWaveCont}
\frac{\partial \rho_{\rm g}}{\partial t}+\frac{\partial{(\rho_{\rm g} v)}}{\partial x} = 0, \ \  
\frac{\partial \rho_{\rm d}}{\partial t}+\frac{\partial{(\rho_{\rm d} u)}}{\partial x} = 0,\ \ 
\end{equation}

\begin{equation}
\label{eq:ShockWaveMotionGas}
\rho_{\rm g} (\frac{\partial v}{\partial t}+v \frac{\partial v}{\partial x}) = - \frac{\partial p}{\partial x} - K(v-u),
\end{equation}

\begin{equation}
\label{eq:ShockWaveMotionDust}
\rho_{\rm d} (\frac{\partial u}{\partial t}+u \frac{\partial u}{\partial x}) =  K(v-u),
\end{equation}

\begin{equation}
\label{eq:ShockWaveEnergyGas}
\rho_{\rm g} (\frac{\partial e}{\partial t} +v\frac{\partial e}{\partial x})=-p\frac{\partial v}{\partial x} + K(u-v)^2,
\end{equation}
where $e$ is the internal energy (temperature) of the gas, which is related to the pressure as
\begin{equation}
\label{eq:EOS}
p=\rho_{\rm g} e (\gamma-1).
\end{equation}

Flow conditions are imposed at the boundaries of the integration interval for the system (\ref{eq:ShockWaveCont})-(\ref{eq:ShockWaveEnergyGas}), and zero initial velocity and discontinuities in the gas pressure, gas
density, and dust density are specified at the initial time. If there is no solid phase in the continuous medium, the analytical solution of this problem is known over the entire region of parameter values. The analytical solution for a dust–gas medium is known for the ``steady-state'' case, that is, for times $t>t_{\rm stop}$ when $t_{\rm stop} \max(u,v) \ll 1$. This solution is obtained from the solution for the gas dynamics by replacing the sound speed in the gas with the sound speed in the gas-dust medium (see, e.g., \cite{LaibePrice2011Test}): 
\begin{equation}
\label{eq:dustysound}
c^*_s=c_s(1+\displaystyle\frac{\rho_{\rm d}}{\rho_{\rm g}})^{-1/2}.
\end{equation}
The propagation velocity of the shock front can be determined over the entire region of parameter values for the one-dimensional problem (see, e.g., \cite{Lehmann2017}). The solution for the shock wave propagation in a two-phase medium in the two-dimensional case can be found in \cite{Golubkina2010,ChertokKurganov2017}. 

We will consider the case when the term on the right-hand side of (\ref{eq:ShockWaveMotionGas}) that corresponds to the drag force appreciably exceeds the term corresponding to the pressure gradient. This makes it possible to neglect the second term in (\ref{eq:ShockWaveEnergyGas}), since it is second order infinitesimal of  $(u-v)$.  

\subsection{Numerical Solution of the DustyShock Problem}
\label{sec:NumDustyShock}

Following the logic of the computational experiments described in Section {\ref{sec:NumDustyWave}}, we will consider a stiflly coupled medium with $K=10^4$ and a high dust concentration $\varepsilon=1$. We considered the initial discontinuities: 
\begin{equation}
\label{SodBoundary1}
 \rho_{L \rm d}=1; \rho_{R \rm d}=0.125;\\
 \rho_{L \rm d}=1; \rho_{R \rm d}=0.125;\\
 \end{equation}
 
 \begin{equation}
\label{SodBoundary2}
 p_L=1; p_R=0.1;\\
 v_L=0; v_R=0; \\
 u_L=0; u_R=0.
\end{equation}
The computations were carried out on a grid with 200 cells in the interval $[0,1]$, with the artificial-viscosity parameter $C2=2$, and with the Courant-Friedrichs-Lewy parameter varied in the range $CFL=0.5 - 0.005$, choosing the time step based on (\ref{eq:tauCFL}). 

\begin{figure*}
  \includegraphics[scale=0.2]{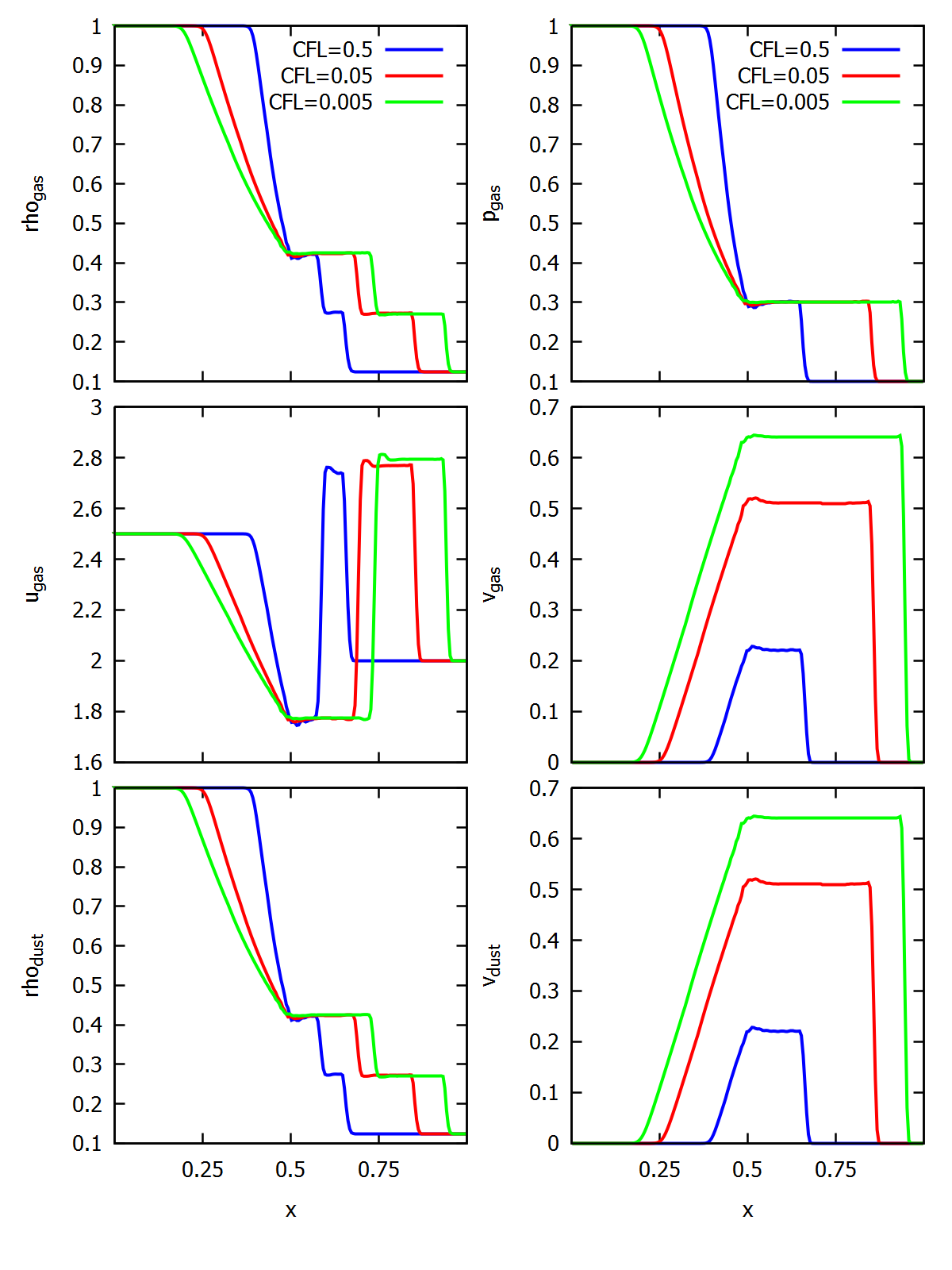} 
  \includegraphics[scale=0.2]{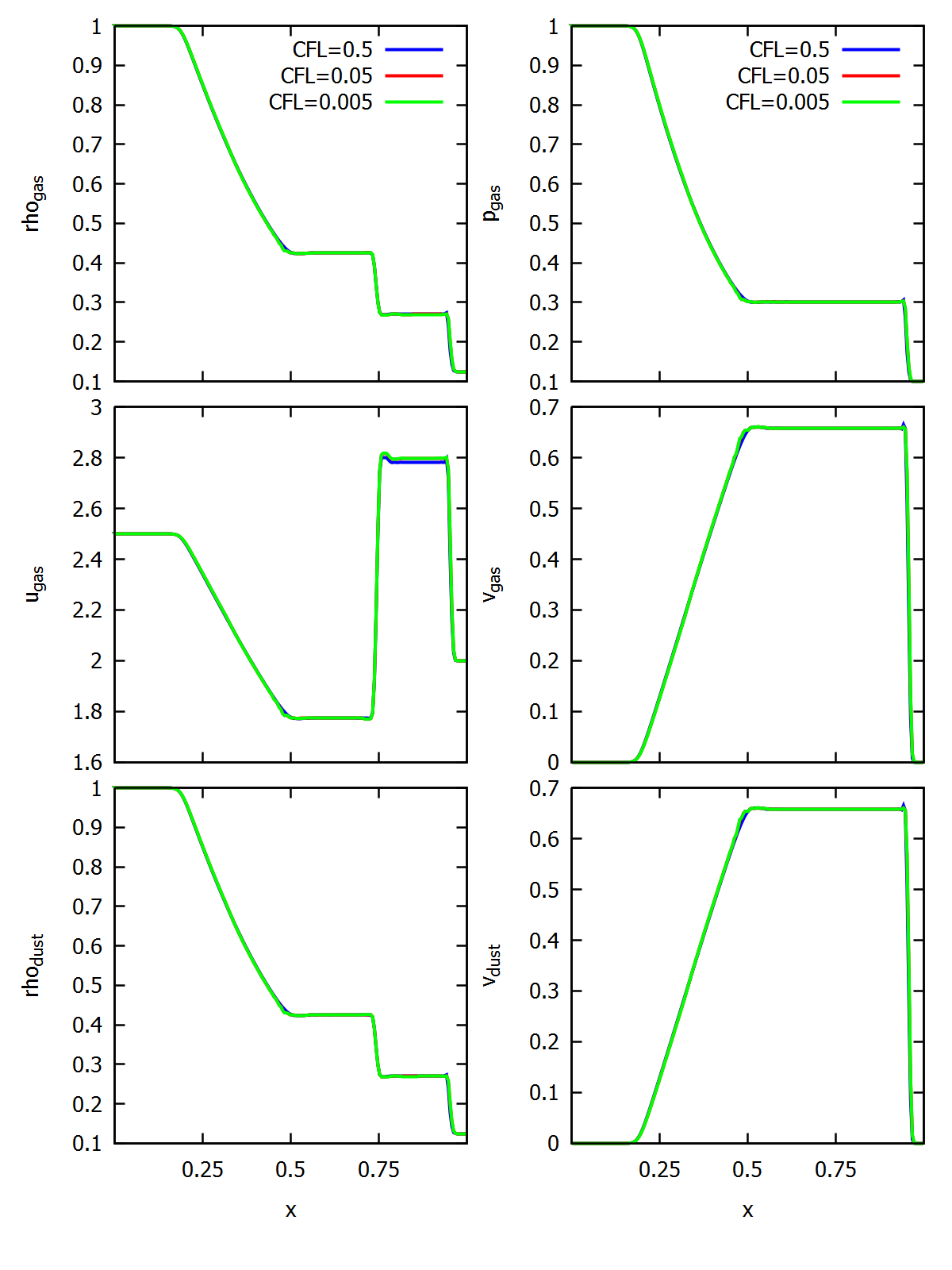}
  \caption{DustyShock solution at time $t=0.37$ obtained with the semi-implicit scheme with operator splitting (\ref{eq:InitFinite}) (left) and the semi-implicit scheme for the barycentric and relative velocities (\ref{eq:Finite})-(\ref{eq:FiniteVU}) (right). Results shown are for the case of a stiffly coupled medium with drag coefficient $K=10000$.}
\label{fig:Shock}
\end{figure*}

It follows from the left six panels of Fig. \ref{fig:Shock} that the scheme (\ref{eq:InitFinite}) with $CFL>0.005$ appreciably underestimates the velocity of the shock and rarefaction wave. The right six panels of Fig. \ref{fig:Shock} show that the finite-difference scheme (\ref{eq:Finite})-(\ref{eq:FiniteVU}) for the barycentric  and relative velocities already yields acceptable computational accuracy for $CFL=0.5$. Similar results were obtained for the scheme based on the analytical solution of (\ref{eq:ExpConst})-(\ref{eq:ExpAll}). 

Analogous to the test of DUSTYWAVE, it follows from these computations that the relation (\ref{eq:sparesSPH}) is not a necessary condition to obtain solutions with acceptable accuracy.

\section{Semi-implicit Scheme for the Mean-weighted and Relative Velocities for a Medium of Gas and Polydisperse Dust}
\label{sec:schemesN}

In this section, we present a method for including several dust fractions in a numerical model for a gas-dust medium, which all exchange momentum with the gas, but not with each other. Results of computations for a medium in which the dust is represented by two fractions with different grain sizes, and consequently different velocity relaxation times $t_{i, {\rm stop}}$ are presented in \cite{Ishiki2017}. The numerical scheme in \cite{Ishiki2017} was based on the analytical solution for a linear system of ordinary differential equations, i.e., on the approach (\ref{eq:ExpConst})-(\ref{eq:ExpAll}). We are not aware of computations of a two-phase medium of gas and $N$ dust fractions with a finely divided solid phase. We have obtained computational formulas for the one-dimensional equations for a system of gas and $N$ dust fractions:

\begin{equation}
\label{eq:Ncontinuum}
\left\{
 \begin{array}{lcl}
       
       \displaystyle\frac{\partial \rho_{\rm g}}{\partial t}+\displaystyle\frac{\partial (\rho_{\rm g} v)}{\partial x}=0, \ \ \ \ 
    
        \displaystyle\frac{\partial \rho_{{\rm d},i}}{\partial t}+\displaystyle\frac{\partial (\rho_{{\rm d},i} u_i)}{\partial x}=0, \ \ i=1,..N,\\
    
        \displaystyle\frac{\partial (\rho_{\rm g} v)}{\partial t} + 
        \displaystyle\frac{\partial (\rho_{\rm g} v^2 )}{\partial x} = 
        f_{\rm g}-\sum_{i} \displaystyle\frac{ \rho_{{\rm d},i}(v-u_i)}{t_{i, {\rm stop}}}, \\
        
        \displaystyle \frac{\partial (\rho_{{\rm d},i} u_i)}{\partial t} + 
        \displaystyle\frac{\partial (\rho_{{\rm d},i} u^2_i) }{\partial x} = 
        f_{{\rm d},i} + \displaystyle\frac{\rho_{{\rm d},i}(v-u_i)}{t_{i, {\rm stop}}}, \ \ i=1,..N.
        
    \end{array}
\right.
\end{equation}

Here, $v$ and $\rho_{\rm g}$ are the velocity and density of the gas, $u_i$ and $\rho_{{\rm d},i}$ the velocity and density of dust fraction $i$, $f_{\rm g}$ the forces acting on the gas other than the drag force, $f_{{\rm d},i}$ the forces acting on dust fraction $i$ apart from the drag force, and $t_{i, {\rm stop}}$ the stopping time of dust fraction $i$.

We can write the equations of motion from (\ref{eq:Ncontinuum}) in the form

\begin{equation}
\label{eq:Nsystem}
\left\{
 \begin{array}{lcl}
    
        \displaystyle 
        \frac{dv}{dt} = a_{\rm g}-\sum_i \varepsilon_i \displaystyle\frac{v-u_i}{t_{i_{\rm stop}}},  \\
        \displaystyle 
        \frac{du_i}{dt} = a_{i} + \displaystyle\frac{v-u_i}{t_{i_{\rm stop}}}, \ \ i=1,..N.
    \end{array}
\right.
\end{equation}

Here, by analogy with the notation of Section \ref{sec:schemes1} $a_{\rm g}$ is the acceleration acting on the gas apart from the drag acceleration, $a_{i}$ the acceleration acting on dust fraction $i$ apart from the drag acceleration, and $\varepsilon_i=\displaystyle\frac{\rho_{{\rm d}, i}}{\rho_{\rm g}}$ the dust of fraction $i$ to gas mass ratio.

We will now turn to a system equivalent to (\ref{eq:Nsystem}) in the variables 
\begin{equation}
\label{eq:newNxy}
y=v+\displaystyle \sum_i \varepsilon_i u_i, \quad x_i=v-u_i,
\end{equation}
where $y$ is the barycentric velocity of the medium and $x_i$ the velocity of dust fraction $i$ relative to the gas:

\begin{equation}
\label{eq:NewNsystem}
\left\{
 \begin{array}{lcl}
        \displaystyle 
        \frac{dy}{dt} = a_{\rm g}+\sum_i \varepsilon_i a_i,  \\
        \displaystyle 
        \frac{dx_i}{dt} = (a_{\rm g}-a_{i}) - \displaystyle\frac{\varepsilon_i+1}{t_{i_{\rm stop}}}x_i-\sum_{j \neq i} \displaystyle\frac{\varepsilon_j}{t_{j_{\rm stop}}}x_j, \ \ i=1,..N.
    \end{array}
\right.
\end{equation}
Note that the inverse transition from the variables $y, x_i$ to the variables $v, u_i$ does not encounter any difficulties (see Appendix \ref{ap:matrix}).

The first equation of the system (\ref{eq:NewNsystem}) can then be solved separately from the remaining equations:
\begin{equation}
\label{eq:yN}
y^{n+1}=y^n+\tau(a_{\rm g}+\sum_i \varepsilon_i a_i).
\end{equation}
Furthermore, using the approach (\ref{eq:Finite}), the numerical solution of the inhomogeneous system of linear equations with dimension $N$ can be found via $O(N^2)$ arithmetic operations:
\begin{equation}
\label{eq:Nsolution}
x^{n+1}_i=-\displaystyle\frac{t_{i_{\rm stop}}}{\tau \varepsilon_i \beta}
\left[ \frac{(1-b_i \beta)(x_i^n+\tau(a_{\rm g}-a_i))}{b_i^2}+
\sum_{j\ne i}\frac{x_j^n+\tau(a_{\rm g}-a_j)}{b_i b_j} \right],
\end{equation}
where 
\begin{equation}
\label{eq:sdet}
\beta=1+\displaystyle\frac{1}{b_1}+\frac{1}{b_2}+...\frac{1}{b_N}, \quad
b_i=\displaystyle\frac{t_{i_{\rm stop}}+\tau}{\varepsilon_i\tau}.
\end{equation}

The velocities of the gas and dust can be determined from the resulting solution (\ref{eq:yN}),(\ref{eq:Nsolution}) of the system  (\ref{eq:NewNsystem}) as follows (see Appendix \ref{ap:polydisperse} for the derivation of these formulas):
\begin{equation}
\label{eq:vel}
v^{n+1}=\displaystyle\frac{y^{n+1}+\sum_i \varepsilon_i x^{n+1}_i}{1+\sum_i \varepsilon_i}, \ \ u^{n+1}_i=\displaystyle\frac{y^{n+1}-(1+\sum_{j \neq i}\varepsilon_j)x^{n+1}_i+\sum_{j \neq i} \varepsilon_j x^{n+1}_j}{1+\sum_i \varepsilon_i}.
\end{equation}

A preliminary discussion of the requirements for the time step $\tau$ when applying the system (\ref{eq:yN})-(\ref{eq:vel}) is presented in Appendix \ref{ap:choose_tau}.

\section{Conclusion}
\label{sec:Resume}

We have presented and compared approaches that are used in modern numerical models applied to the computation of the dynamics of gas-dust circumstellar disks. Circumstellar disks consist of gas and solid bodies ranging from submicron dust to meter-sized bodies. The stopping time for small dust grains is much less than the dynamical time scale; i.e., it
comprises a small fraction of an orbit of the disk around the central protostar. We have limited our consideration to models in which the gas and dust are treated like interpenetrating continuous media that can exchange momentum. We have focused on the suitability of various methods for modeling a medium with (a) arbitrary (short or long) stopping times for the dust and (b) arbitrary dust concentration in the gas (the dust to gas mass ratio varied from 0.01 to 1). We will now summarize our main results.

The method for computing the momentum exchange based on analytical solutions have infinite-order accuracy in time. Applying this method to a two-phase medium makes it possible to satisfy conditions (a) and (b) with minimal computational costs making it an optimal approach for the solution of non-stationary problems.

The semi-implicit scheme with operator splitting applied to a two-phase medium is stable, that is, it enables computations with $\tau>t_{\rm stop}$, but its actual accuracy becomes unacceptably low for time steps $\tau>\displaystyle\frac{t_{\rm stop}}{\varepsilon}$. This scheme is not recommended, since it requires time steps as small as an explicit scheme when $t_{\rm stop} \ll 1$ and $\varepsilon \approx 1$. 

Our proposed semi-implicit scheme for the barycentric and relative velocities with first-order accuracy in time for a two-phase medium enables satisfaction of the conditions (a) and (b) with computational costs comparable to the method with infinite-order accuracy. An advantage of this method is that it can be extended to a regime where the dust is polydisperse, in other words, it is represented by several dust fractions with different stopping times. We have presented formulas for the computation of the gas and dust velocities when the dust has $N$ fractions, each of which exchanges momentum with the gas; these formulas require $O(N^2)$ arithmetic operations at each time step.

\section*{Acknowledgements}

We thank Ya. N. Pavlyuchenkov for detailed discussions of this work, which was supported by the Russian Science Foundation (grant 17-12-01168).

\appendix

\section{Transition matrix}
\label{ap:matrix}
The matrix for the transition from the velocities of the individual components of the medium to the barycentric and relative velocities (\ref{eq:newNxy}) has a special form and is analytically reversible:
\begin{equation*}
\begin{pmatrix}
1 & \varepsilon_1 & \varepsilon_2 & \varepsilon_3 & ... & \varepsilon_n \\
1 & -1            & 0             & 0             & ... & 0 \\
1 & 0             & -1            & 0             & ... & 0 \\
... & ...         & ...           & ...           &...  &...\\
1 & 0             & 0             & 0             & 0   & -1
\end{pmatrix}^{-1}=
\end{equation*}

\begin{equation}
\displaystyle\frac{1}{1+\displaystyle \sum_i \varepsilon_i}
\begin{pmatrix}
1 & \varepsilon_1 & \varepsilon_2 & \varepsilon_3 & ... & \varepsilon_n \\
1 & -(1+\displaystyle \sum_{i \ne 1} \varepsilon_i)     & \varepsilon_2 & \varepsilon_3 & ... & \varepsilon_n \\
1 & \varepsilon_1 & -(1+\displaystyle \sum_{i \ne 2} \varepsilon_i)             & \varepsilon_3 & ... & \varepsilon_n \\
... & ...         & ...           & ...           & ... &...\\
1& \varepsilon_1  & \varepsilon_2 & \varepsilon_3 & ... & -(1+\displaystyle \sum_{i \ne n} \varepsilon_i)
\end{pmatrix},
\end{equation}
where the eigenvalues of the matrix $\lambda_1=-1$ have multiplicity $(n-1)$ and $\lambda_{2,3}= \pm \sqrt {1+\displaystyle \sum_i \varepsilon_i}$ making it possible to estimate the condition number of the matrix. In particular, if $\displaystyle \sum_i \varepsilon_i < 1$, the matrix is well conditioned.

\section{Derivation of Computational Formulas for Polydisperse Dust}
\label{ap:polydisperse}
The subsystem of the equations for the relative velocities from (\ref{eq:NewNsystem}) has the form
\begin{equation}
\label{eq:general_matrix_eq}
\displaystyle\frac{d\bf {x}}{dt}=- T \bf x + \bf q,
\end{equation}
where 
\begin{equation}
T=
\begin{pmatrix}
\displaystyle\frac{\varepsilon_1+1}{t_{1_{\rm stop}}}& \displaystyle\frac{\varepsilon_2}{t_{2_{\rm stop}}} & ... & \displaystyle\frac{\varepsilon_N}{t_{N_{\rm stop}}} \\
\displaystyle\frac{\varepsilon_1}{t_{1_{\rm stop}}} & \displaystyle\frac{\varepsilon_2+1}{t_{2_{\rm stop}}} & ... & \displaystyle\frac{\varepsilon_N}{t_{N_{\rm stop}}} \\
... & ...                                    & ... &... \\
\displaystyle\frac{\varepsilon_1}{t_{1_{\rm stop}}}   & 
\displaystyle\frac{\varepsilon_2}{t_{2_{\rm stop}}}                                   & ...   
& \displaystyle\frac{\varepsilon_N+1}{t_{N_{\rm stop}}}
\end{pmatrix},
\quad
\bf x = 
\begin{pmatrix}
x_1\\
x_2\\
...\\
x_N
\end{pmatrix},
\quad
\bf q=
\begin{pmatrix}
a_{\rm g}-a_1\\
a_{\rm g}-a_2\\
...\\
a_{\rm g}-a_N
\end{pmatrix}.
\end{equation}

We can write an approximation of Eq. (\ref{eq:general_matrix_eq}) using the implicit first-order method:

\begin{equation}
\displaystyle\frac{\textbf{x}^{n+1} - \textbf{x}^n }{\tau}= - T \textbf{x}^{n+1}+ \textbf q,
\end{equation}
which yields the system of linear equations for $\textbf{x}^{n+1}$:

\begin{equation}
\label{eq:ngrain_linear}
(I+\tau T) \textbf{x}^{n+1}=\textbf{x}^n+\tau \textbf{q}.
\end{equation}

In Eq. (\ref{eq:ngrain_linear}), the matrix $(I+\tau T)$ can be transformed into a special form that is convenient for finding the inverse matrix through the substitution
\begin{equation}
\label{eq:xtoz}
z_i=\displaystyle\frac{\tau \varepsilon_i}{t_{i_{\rm stop}}}x_i^{n+1} \quad 
b_i=\displaystyle\frac{t_{i_{\rm stop}}+\tau}{\varepsilon_i\tau}.
\end{equation}
Equation (\ref{eq:ngrain_linear}) is then equivalent to
\begin{equation}
B\textbf{z}=\textbf{x}^n+\tau \textbf{q}, 
\quad
B=
\begin{pmatrix}
1+b_1 & 1 & ... & 1 \\
1 & 1+b_2 & ... & 1 \\
... & ... & ... &... \\
1 &  1    & ... & 1+b_N
\end{pmatrix}.
\quad
\end{equation}
Setting
\begin{equation}
\beta=1+\displaystyle\frac{1}{b_1}+\frac{1}{b_2}+...\frac{1}{b_N},
\end{equation}
it is straightforward to see that
\begin{equation}
B^{-1}=-\displaystyle\frac{1}{\beta}
\begin{pmatrix}
\displaystyle\frac{1-b_1 \beta}{b_1^2} & \displaystyle\frac{1}{b_1 b_2} & ... & \displaystyle\frac{1}{b_1 b_N} \\
\displaystyle\frac{1}{b_2 b_1} & \displaystyle\frac{1-b_2 \beta}{b_2^2} & ... & \displaystyle\frac{1}{b_2 b_N} \\
... & ... & ... &... \\
\displaystyle\frac{1}{b_N b_1} &  \displaystyle\frac{1}{b_N b_2}    & ... & 
\displaystyle\frac{1-b_N \beta}{b_N^2}
\end{pmatrix}.
\quad
\end{equation}
We can then write expressions for the components of the vector $\textbf{x}^{n+1}$:
\begin{equation}
%\label{eq:Nsolution}
x^{n+1}_i=-\displaystyle\frac{t_{i_{\rm stop}}}{\tau \varepsilon_i \beta}
\left[ \frac{(1-b_i \beta)(x_i^n+\tau(a_{\rm g}-a_i))}{b_i^2}+
\sum_{j\ne i}\frac{x_j^n+\tau(a_{\rm g}-a_j)}{b_i b_j} \right]
\end{equation}

\section{Conditions on Choice of the Time Step for the Computational Scheme for Polydisperse Dust-preliminary Results}
\label{ap:choose_tau}

We can derive from (\ref{eq:ngrain_linear}) sufficient conditions under which $\textbf{x}^{n+1}$ will depend continuously on $\textbf{x}^n+\tau \textbf{q}$. If the matrix $B$ had diagonal predominance, that is,
for all $i$ 

\begin{equation}
\label{eq:tau_condition}
1+\displaystyle\frac{t_{i_{\rm stop}}+\tau}{\varepsilon_i\tau} \geq N-1,
\end{equation}
such a continuous dependence is attained. In particular, if for all $i$
\begin{equation}
\label{eq:easycaseTau}
(N-2) \varepsilon_i<1,
\end{equation}
there is no restriction on the time step $\tau$; otherwise, a sufficient condition for correctness of the numerical scheme for all $i$ for which (\ref{eq:easycaseTau}) is violated is 
\begin{equation}
\label{eq:tau_suff}
\tau \leq \displaystyle \frac{ t_{i_{\rm stop}}}{(N-2) \varepsilon_i-1}.
\end{equation} 

Note that the condition (\ref{eq:easycaseTau}) includes $N$ - -the number of beams, into which the entire dust subdisk is divided. The optimal separation of the particles according to their sizes (masses) from the point of view of describing the growth of the dust is discussed, for example, in \cite{Drazkowska2014,Ohtsuki}. Therefore, we wish to elucidate whether cases arise in simulations of disks when $\varepsilon_i$ decreases more slowly than $N$ grows when the range of dust size in the beam is decreased. Furthermore, in some specific cases, the condition (\ref{eq:tau_suff}) can be fairly strict if the solid phase is represented by a large number of components; therefore, further study is required to address the question of whether this condition is necessary.

\bibliographystyle{unsrt}
\bibliography{Stoyanovskaya}

\end{document}